# Uniqueness theorem for the non-local ionization source in glow discharge and hollow cathode

*V.V.Gorin*

Kyiv National Taras Shevchenko University,

Moscow Institute of Physics and Technology

**Abstract.** The paper is devoted to the proof of the uniqueness theorem for solution of the equation for the non-local ionization source in a glow discharge and a hollow cathode in general 3D geometry. The theorem is applied to wide class of electric field configurations, and to the walls of discharge volume, which have a property of incomplete absorption of the electrons. Cathode is regarded as interior singular source, which is placed arbitrarily close to the wall. The existence of solution is considered also. During the proof of the theorem many of useful structure formulae are obtained. Elements of the proof structure, which have arisen, are found to have physical sense. It makes clear physical construction of non-local electron avalanche, which builds a source of ionization in glow discharge at low pressures. Last has decisive significance to understand the hollow cathode discharge configuration and the hollow cathode effect.

## 1. Introduction

The problem of creation of the hollow cathode theory, - the theory for a glow discharge device in gases, which was invented by Paschen almost yet hundred years ago [1], producing anomalously high currents at the same voltages of discharge compared the glow discharge devices, which have no geometry of hollow cathode, - considerably stipulated for non-local ionization. The classical theory of the Engel and Shteenbeck cathode sheath [2] for glow discharge in simple geometry used the Townsend formula for a source of ionization [3], which had in mind local dependence of ionization on the electric field. The local (or two-fluid) models of a glow discharge [4] gave qualitatively true description of the electric field and the electron and ion current density distributions in the cathode sheath of the plane capacitor at low current densities and not too low pressures. However they couldn't catch really the specific of negative glow. Last has low electric field, and the ionization here is produced with electrons gained the energy in another place, namely in the high field of the cathode sheath. Therefore the region of negative glow in the local model could not be obtained in principle, and the cathode sheath here is always contacting with a positive column. For a hollow cathode the model with local ionization turned out unacceptable at all, because both cathode sheath and plasma of discharge have significantly non-local properties.

On the route to the non-local ionization theory there were developed hybrid models - instead of two-fluid models, - in which electrons produced ionization were considered apart from slow electrons, latter providing balance of current and electric charge [5]. In a hybrid model ions and slow electrons are described in terms of drift and diffusion, but fast electrons are described with the aid of the boltzmann equations or simulated with Monte-Carlo methods [6 - 8]. It was shown [9] that hybrid models describe a density of plasma in a glow discharge much better, though they are more complicated.

Hybrid models set a problem of description of non-local ionization source, for which the Townsend formula is not available. The boltzmann equation for fast electrons is many-dimensional, which makes difficulties in its use in calculations. The Monte-Carlo method is in essence a computing experiment, which supplies empirical data about numerical simulation of ionization source without understand reasons of results obtained.

The author of present paper had managed earlier to simplify this problem by use of original (not the hilbert [10]) averaging of the boltzmann equation for fast electrons, in result it was obtained the non-local equation for a source of ionization in glow discharge and hollow cathode [11, 12]:

$$s(\mathbf{r}) = \int_{\partial\Omega} d^2 r' G_0(\mathbf{r}, \mathbf{r}') j_n(\mathbf{r}') + \int_{\Omega} d^3 r' G_0(\mathbf{r}, \mathbf{r}') s(\mathbf{r}').$$

Here $s(\mathbf{r})$ is a density of ionization source, $j_n$ is an electron flow density from the cathode (really first summand gives a contribution only in the part of the boundary $\partial\Omega$ of the spatial area $\Omega$ of the glow discharge, in which electrons are incoming from external source - a cathode), the definition of $G_0(\mathbf{r}, \mathbf{r}')$ see below. First summand in the right hand side of the equation describes an ionization with cathode electrons only, second summand describes an ionization with secondary electrons arisen from impact ionization inside discharge volume. The non-local source equation has mathematical class of an integral Fredholm second kind equation. It does not conflict with a local model of the Townsend ionization source, but rather is a non-trivial and far-reaching generalization. Namely: if one rewrites the non-local equation in terms of one-dimensional spatial model of plane capacitor (without hollow cathode configuration) and simplifies the electron kinetics by rejecting elastic scattering and discreteness of energy losses in inelastic processes, the equation gets a class of an integral second kind Volterra equation [13, 14]. In a case, when the non-local effects can be neglected, the kernel of equation quits to depend on its second argument. This kernel degeneration enables to transform the integral equation into a differential equation, which coincides literally with the Townsend equation $dj_e/dx = \alpha(x)j_e$.

Present paper is devoted to the proof of uniqueness for solution of non-local equation if it exists. Though there exists a set of theorems for existence and uniqueness of solutions in the theory of the Fredholm equations [15 - 17], one cannot use them because: 1) the kernel is defined here not evidently, but as a solution of the linear boltzmann equation with differential and integral operators, the parameters of which have rather general physical properties; 2) the domain of kernel definition is defined implicitly also, its geometry is varied in wide range of glow discharge devices, a hollow cathode of arbitrary shape is one of possibilities; 3) the integral term of the equation is usually not of low value in comparison with absolute term (secondary electrons usually contribute more into ionization against primary, cathode, ones); 4) the kernel is not hermitian (or symmetric) one.

The question about existence of a solution is tied with a question about existence of kernel of the equation, the answer depends on existence of fundamental solutions of the boundary problem for a *stationary* boltzmann equation for fast electrons. Not all configurations of the electrical field and not all kinds of boundary conditions guarantee an existence of a fundamental solution for stationary problem. For example, if electric field is equal to zero in some region of ionization one cannot neglect initial velocity of secondary electrons, as it was done in derivation of the equation, because electrons would be accumulated with no limits in this area, consequently a solution could not be stationary. Necessary conditions for existence of fundamental solution for auxiliary differential operator of the problem come to the Fredholm alternatives [15]. Investigation of *sufficient* conditions for existence of solution is not easy problem, and in the paper we make the assumption that necessary fundamental solution for auxiliary operator exists.

The proof is divided on five lemmas and final proving of the theorem. In lemma 1 it is proved uniqueness of zero solution for homogeneous differential equation, which is generated by auxiliary differential operator. This operator defines the distribution function for "fast" electrons, which are fortunate not suffer any inelastic scattering as long as they appear from a source of ionization (or from the cathode). Lemma 2 formulates analogues statement for conjugate operator. From these results it follows the uniqueness of fundamental solution for auxiliary differential operator (if it exists), so the lemma 3 is devoted to prove this. In lemma 4 the fundamental solution of the boltzmann equation and some its properties are constructed from the solution and properties of fundamental solution for auxiliary operator. On the ground of properties obtained it is proved the uniqueness of definition and the nilpotency of the kernel of the integral equation, shown above. Namely the property of nilpotency (vanishing of some power of appropriate operator) gives here a possibility to state a convergence of the Neumann series and obtain the formula for solution of the integral equation, which is final part of the theorem proof.

Thus, a set of useful formulae is obtained during the proof of the theorem. The constructions arisen have clear physical sense.

## 2. Definitions and properties of physical values

Consider the domain $\Xi:(\mathbf{r},\mathbf{v})$, $\mathbf{r}\in\Omega$, $\mathbf{v}\in R^3$, $\Xi=\Omega\times R^3$, $\Omega\subset R^3$ of 6-dimensional phase space for mechanics of electron motion, $\Omega$ is its spatial domain - open connected set in $R^3$, - the bound $\partial\Omega$ of which consists of piecewise smooth surfaces and is defined with the geometry of the glow discharge device.

$\omega_{ion}(v)$ is an average electron impact ionization rate at the electron velocity $v$, it is nonnegative continuous function, which has an energy threshold:

$$\omega_{ion}(v)=0 \text{ at } \frac{m_e v^2}{2e}\leq\varepsilon_{ion}, \tag{1}$$

here $e$ is an elementary charge, $\varepsilon_{ion}$ is the electron impact ionization energy of the atom, eV, $m_e$ is a mass of electron.

The operator of elastic scattering

$$L_{el}(\mathbf{v})=\omega_{el}(v)\sum_{i=1}^{3}\sum_{k=1}^{3}\frac{\partial}{\partial v_i}\left(\delta_{ik}v^2-v_i v_k\right)\frac{\partial}{\partial v_k}, \quad \omega_{el}(v)\geq 0 \tag{2}$$

conserves a number of electrons and kinetic energy of scattered electrons $w=m_e v^2/(2e)$, eV, here $\omega_{el}(v)$ is an average rate of elastic electron scattering on (almost) immovable gas atoms at the electron velocity $v$. It is continuous function, positive at all $v > 0$.

The operator of inelastic scattering at electron impact excitation and ionization

$$L_{in}(\mathbf{v})f(\mathbf{v}) = \int d^3v' \mu(\mathbf{v},\mathbf{v}')f(\mathbf{v}') - \omega(v)f(\mathbf{v}), \quad \mu(\mathbf{v},\mathbf{v}') = \mu_3\left(v,v',\frac{\mathbf{v}\cdot\mathbf{v}'}{vv'}\right) \quad (3)$$

conserves a number of electrons involved into inelastic scattering with atoms (here we mind *primary* electrons, - we do not include *secondary* electrons, which appear in ionization in addition):

$$\omega(v) = \int d^3v' \mu(\mathbf{v}',\mathbf{v}) = 2\pi \int_0^\infty v'^2 dv' \int_0^\pi d\theta \sin\theta \, \mu_3(v',v,\cos\theta), \quad (4)$$

$\mu(\mathbf{v},\mathbf{v}') \geq 0$, in general case it is generalized function, the kernel of linear operator $\hat{\mu}: C(R^3) \Rightarrow C(R^3)$, which, by its physical sense of average rate of inelastic processes transforming the velocity of electron from value $\mathbf{v}'$ (second argument) to infinitesimal domain, neighborhood of $\mathbf{v}$, have to be a continuous operator on a class of continuous finite functions. The property of the energy dissipation in inelastic processes puts a restriction

$$\mu(\mathbf{v},\mathbf{v}') = 0 \text{ at } \frac{m_e\mathbf{v}^2}{2e} + \varepsilon_{in} \geq \frac{m_e\mathbf{v}'^2}{2e}, \quad \varepsilon_{in} > 0. \quad (5)$$

$\varepsilon_{in}$ is the smallest of the energy thresholds for inelastic processes, eV. On physical reasons $\varepsilon_{ion} > \varepsilon_{in}$ because atomic ionization energy always exceeding excitation energy of any level from discrete energy spectrum.

It follows from restriction (5) that integration in (4) could be narrowed to the ball

$$\omega(v) = \int_{\frac{m_e\mathbf{v}'^2}{2e} < \frac{m_e\mathbf{v}^2}{2e} - \varepsilon_{in}} d^3v' \, \mu(\mathbf{v}',\mathbf{v}) = 2\pi \int_0^{V(v)} v'^2 dv' \int_0^\pi d\theta \sin\theta \, \mu_3(v',v,\cos\theta),$$

$$V(v) = \sqrt{v^2 - \frac{2e}{m_e}\varepsilon_{in}},$$

from which follows that the average rate $\omega(v)$ of inelastic processes for electron with velocity $\mathbf{v}$ has lower energy threshold $\varepsilon_{in}$, namely, $\omega(v) = 0$ at $\frac{m_e v^2}{2e} \leq \varepsilon_{in}$. We consider the function $\omega(v)$ be continuous.

$\mathbf{E}(\mathbf{r}) = -\nabla\varphi(\mathbf{r})$ is an electric field defined in the domain $\Omega$. Choose the reference level of electric potential $\varphi(\mathbf{r})$ so that it would have non-positive values only, and maximum of potential value in the closing $\overline{\Omega}$ would be equal to zero. Then the minimal value $\min_{\mathbf{r}\in\Omega}\{\varphi(\mathbf{r})\}=-U$. $U$ usually coincides with a voltage of discharge.

Let us define, that *the local maximum* of electric potential on the closing $\overline{\Omega}$ is non-empty closed connected set $M$: $\mathbf{r} \in M$, $M \subset \overline{\Omega}$ having properties: if the point $\mathbf{r}_0 \in M$ has chosen, then 1) for any other point $\mathbf{r} \in M$ (if it exists) the equality $\varphi(\mathbf{r}) = \varphi(\mathbf{r}_0)$ is true, 2) for any $\delta > 0$ there exists a $\delta$-neighborhood $\delta_M$: $\forall \mathbf{r}' \in \delta_M$: $\exists \mathbf{r} \in M$: $|\mathbf{r}' - \mathbf{r}| < \delta$, $\varphi(\mathbf{r}') < \varphi(\mathbf{r}_0)$. Here generality quantifier $\forall$ designates "for any", existence quantifier $\exists$ designates "there exists". The *local minimum* we define in analogy, by changing sign "less" (<) to sign "more" (>) in last inequality.

Because of the local maximum of the potential is a local minimum of the electron potential energy, and inelastic processes occur only at electron kinetic energy exceeding the threshold value $\varepsilon_{in}$, it might exist closed domain with slow electrons in the neighborhood of local maximum of potential at sufficiently small electron velocity, the electrons here do not participate inelastic processes, and thus its total mechanical energy conserves. All such domains of electron phase space (if they exist) let us join into a zone of slow electrons:

Define *a zone of slow electrons:*

$$\Xi_s = \Xi_s(\varphi, \varepsilon_{in}) = \overline{\Xi} \cap \bigcup_{p=1}^{P}\left\{-\varphi\left(\mathbf{r}_0^{(p)}\right) \leq \varepsilon(\mathbf{r},\mathbf{v}) \leq \varepsilon_p\right\}, \quad \varepsilon_p = \min[\delta\varphi_p, \varepsilon_{in}] - \varphi\left(\mathbf{r}_0^{(p)}\right),$$

$$\varepsilon(\mathbf{r},\mathbf{v}) = \frac{m_e v^2}{2e} - \varphi(\mathbf{r}),$$

here $\mathbf{r}_0^{(p)}$ is a point of local maximum number $p$ of potential, $\delta\varphi_p$ is a trap depth, index $p = 1,..,P$ lists mutually insolated areas where the inequalities for small kinetic energy are true (slow electron cannot penetrate from one area to another through potential barrier).

Define *a zone of "fast" electrons* as $\Xi_{in} = \Xi \setminus \Xi_s$. The set $\Xi_{in}$ is an open connected set, so it is a domain. Openness is obvious, connectedness is provided with unrestrictedness of the energy from the top in $\Xi_{in}$.

Any two points $(\mathbf{r}_1, \mathbf{v}_1) \in \Xi_{in}, (\mathbf{r}_2, \mathbf{v}_2) \in \Xi_{in}$ could be connected with piecewise smooth curve inside $\Xi_{in}$ as follows: straights $C_1: \{(\mathbf{r}_1, \mathbf{v}_1 t), 1 \leq t < \infty\}$, $C_2: \{(\mathbf{r}_2, \mathbf{v}_2 t), 1 \leq t < \infty\}$ belong to $\Xi_{in}$, because the increase of kinetic energy in the point concerned cannot transform "fast" electron into slow one. Let be $U = \max_{\mathbf{r}\in\Omega}\{-\varphi(\mathbf{r})\}$. The hypersurface $\varepsilon(\mathbf{r},\mathbf{v}) = \max\{U + 2\varepsilon_{in}, \varepsilon(\mathbf{r}_1, \mathbf{v}_1), \varepsilon(\mathbf{r}_2, \mathbf{v}_2)\}$ belongs to $\Xi_{in}$ at all $\mathbf{r} \in \Omega$, because the kinetic energy in it is not less than $2\varepsilon_{in}$ everywhere. Connectedness of this hypersurface is provided by connectedness of $\Omega$. The straights $C_1$ and $C_2$ cross the hypersurface given, because total energy is unrestricted along them. So, one can connect the points of crossing the hypersurface with third curve because of hypersurface connectedness.

Besides $\Xi_{in}$, we define its subsets $\Xi_{in+\delta\varepsilon} = \Xi \setminus \Xi_s(\varphi, \varepsilon_{in} + \delta\varepsilon)$, which distinguish from $\Xi_{in}$ by that we use overstated value of the threshold $\varepsilon_{in} + \delta\varepsilon$, $\delta\varepsilon > 0$ in the definition of $\Xi_s$ (and then $\Xi_{in}$) against former threshold $\varepsilon_{in}$. In particular, *the domain of ionizing electrons* $\Xi_{ion} = \Xi \setminus \Xi_s(\varphi, \varepsilon_{ion})$ we define similarly the domain of "fast" electrons, a single difference is that minimal energy of inelastic processes $\varepsilon_{in}$ in the definition we substitute with ionization energy $\varepsilon_{ion}$. Obviously $\Xi_{ion} \subset \Xi_{in}$, $\overline{\Xi}_{ion} \subset \overline{\Xi}_{in}$. Besides, the subsets of $\Xi_{in}$, which are restricted from the top by energy, we designate as $\Xi_{in+\delta\varepsilon}^{E} = \Xi_{in+\delta\varepsilon} \cap \{\varepsilon(\mathbf{r}, \mathbf{v}) < E\}$.

The subset of 3D velocity space

$$\mathbf{v} \in S\Xi_{in}(\mathbf{r}): \quad (\mathbf{r}, \mathbf{v}) \in \Xi_{in} \cap \{\mathbf{r} = \text{const}\}$$

let be named *a local section* of domain $\Xi_{in}$. A local section of domain $\Xi_{in}$ either is coincident all velocity space $R^3$ (if $-\varphi(\mathbf{r}) > \varepsilon_{in}$), or it is an exterior of closed ball with velocity radius, which does not exceeding $\sqrt{2e\varepsilon_{in}/m_e}$ (slow electrons from the $\Xi_s$ are distributed inside the ball).

Let us define *natural directions* in the point of 6D phase space (excluding singular points $(\mathbf{r}, \mathbf{v}): \{\mathbf{v} = 0\} \cap \{\mathbf{E}(\mathbf{r}) = 0\}$). Their "naturalness" is defined with properties of the operator of the boltzmann equation, see below.

1) a normal to the hypersurface of energy

$$\mathbf{N}_\varepsilon = \left(\frac{\partial \varepsilon}{\partial \mathbf{r}}, \frac{\partial \varepsilon}{\partial \mathbf{v}}\right) = \left(\mathbf{E}, \frac{m_e}{e}\mathbf{v}\right),$$

2) a direction of phase flow (phase trajectories of collision-free electron motion)

$$\mathbf{N}_t = \left(\mathbf{v}, -\frac{e}{m_e}\mathbf{E}\right),$$

3) a spatial-like direction having spatial component, which is orthogonal to the velocity and electric field

$$\mathbf{N}_{r\perp} = (\mathbf{v} \times \mathbf{E}, \mathbf{0}),$$

4) a velocity-like direction, having velocity component, which is orthogonal to the velocity and electric field

$$N_{v\perp} = (0, v \times E),$$

5) a velocity-like direction, which is orthogonal to $N_{v\perp}$, having a velocity component, which is orthogonal to velocity

$$N_{vE} = \left(0, E - v\frac{E \cdot v}{v^2}\right),$$

6) a direction, which is orthogonal to five ones above

$$N_5 = \left(-\left(\frac{m_e}{e}v^4 + \frac{e}{m_e}(E \cdot v)^2\right)E + (E \cdot v)\left(\frac{m_e}{e}v^2 + \frac{e}{m_e}E^2\right)v, (E \times v)^2 v\right).$$

In the case when $v \times E = 0$ the directions 3), 4), 5) and 6) we define with the use of arbitrary two vectors $e_1, e_2$ of 3-dimensional Euclidean space, which constitute the orthogonal triad with non-zero $v$ or $E$:

$$N_{r\perp} = (e_1, 0), \ N_{v\perp} = (0, e_1), \ N_{vE} = (0, e_2), \ N_5 = (e_2, 0).$$

So defined directions do not consist mutually orthogonal set of directions in 6D space (because a scalar product: $N_t \cdot N_{vE} = -\frac{e}{m_e v^2}(E \times v)^2$ is not equal to zero in general case), but they consist a complete basis of directions in (regular) point $(r, v) \in \Xi_{in}$.

Indeed, first four 6D-vectors: $N_\varepsilon, N_t, N_{r\perp}, N_{v\perp}$ are mutually orthogonal (in 6D Euclidean space). If a set of vectors

$$N_\varepsilon, N_t, N_{r\perp}, N_{v\perp}; N_{vE}$$

is linear dependent, then equality $N_{vE} = aN_\varepsilon + bN_t + cN_{r\perp} + dN_{v\perp}$ must be true at some values of factors $a, b, c, d$. Taking scalar products $N_{vE} \cdot N_\varepsilon$, $N_{vE} \cdot N_t$, $N_{vE} \cdot N_{r\perp}$, $N_{vE} \cdot N_{v\perp}$ one can find, that only factor $b$ can be non-zero: $N_{vE} = bN_t$. But it is obvious from definition, that vectors $N_{vE}, N_t$ are not collinear. This contradiction means, that the set $N_\varepsilon, N_t, N_{r\perp}, N_{v\perp}; N_{vE}$ is linear independent.

Let us define *a function class* $f(r, v) \in D(\Xi_{in})$. The functions are simultaneously

1) finite functions, support of which is belonging to closing $\overline{\Xi}_{in}$ of the domain $\Xi_{in}$;

2) continuously differentiable along direction $N_t$ of the phase flow;

3) twice continuously differentiable on $v$ in local section $v \in S\Xi_{in}(r)$ along sphere $|v| = \text{const}$;

4) having its continuous extension and continuous extension of all listed derivatives into the piece of boundary $\partial \Xi_{in} \cap \{\partial \Omega \times R^3\}$ of domain $\Xi_{in}$;

5) as the problem has no differential operators along the rest of three directions $N_\varepsilon$, $N_{r\perp}$ and $N_5$, we guess necessary the continuity of function $f$ in $\Xi_{in}$ itself, nothing about its derivatives in these directions.

In analogy we define the function classes $D(\Xi_{in+\delta\varepsilon})$, $D(\Xi_{in+\delta\varepsilon}^E)$ and so on, having in mind other domains of definitions in the item 1).

Let us designate as $C(\Xi_{in})$, $C(\Xi_{in+\delta\varepsilon})$, $C(\Xi_{in+\delta\varepsilon}^E)$ classes of finite functions, which are continuous on closings of domains of their definition.

The operator of electron scattering on atoms of gas

$$L(\mathbf{v}) = L_{el}(\mathbf{v}) + L_{in}(\mathbf{v}), \tag{6}$$

also operator of phase flow

$$\mathbf{v}\cdot\frac{\partial}{\partial\mathbf{r}} - \frac{e}{m_e}\mathbf{E}(\mathbf{r})\cdot\frac{\partial}{\partial\mathbf{v}}$$

for collision-free motion of electron in the electric field can be considered as acting from $D(\Xi_{in})$ to $C(\Xi_{in})$.

Define the generalized function $g(\mathbf{r},\mathbf{v};\mathbf{r}',\mathbf{v}')$, $(\mathbf{r},\mathbf{v};\mathbf{r}',\mathbf{v}') \in \Xi_{in}\times\Xi_{in}$ as a fundamental solution of stationary boltzmann equation

$$\left(\mathbf{v}\cdot\frac{\partial}{\partial\mathbf{r}} - \frac{e}{m_e}\mathbf{E}(\mathbf{r})\cdot\frac{\partial}{\partial\mathbf{v}} - L(\mathbf{v})\right)g = \delta^3(\mathbf{r}-\mathbf{r}')\delta^3(\mathbf{v}-\mathbf{v}') \tag{7}$$

in the domain $(\mathbf{r},\mathbf{v}) \in \Xi_{in}$ with a point source in the right hand side, and satisfying the *condition of incomplete absorption* in the boundary $(\partial\Omega\times R^3)\cap\overline{\Xi}_{in}$ of the domain $\Xi_{in}$:

$$\mathbf{r}\in\partial\Omega,\ \mathbf{v}\cdot\mathbf{n}_{\partial\Omega}\leq 0:\ g(\mathbf{r},\mathbf{v};\mathbf{r}',\mathbf{v}') = \beta\left(v,\frac{\mathbf{v}}{v}\cdot\mathbf{n}_{\partial\Omega}\right)g(\mathbf{r},\mathbf{v}-2(\mathbf{v}\cdot\mathbf{n}_{\partial\Omega})\mathbf{n}_{\partial\Omega};\mathbf{r}',\mathbf{v}'), \tag{8}$$

here $\mathbf{n}_{\partial\Omega}$ is an external unit 3D normal in the boundary $\partial\Omega$ of the domain $\Omega$, $\beta: 0\leq\beta\leq 1$ is the reflection factor for electrons in the boundary of discharge.

Physical sense of the condition of incomplete absorption is: the boundary $\partial\Omega$ of discharge volume $\Omega$ absorbs the part $1-\beta$ of all electrons in average, which get it, the rest of electrons suffer elastic ("mirror") reflection by the boundary. In this way, the cathode - a boundary emitter of electrons - is convenient to simulate as a given singular source, which is located inside $\Omega$, but arbitrarily close to its real boundary position. This convenience can be explained by active feature of cathode as electron emitter, and passive feature of anode and walls as electron absorbers and reflectors: the flow of cathode emission must be given as input data of the problem, but flows of absorption and reflection can be found only in result of problem solving.

The subset of functions from $D(\Xi_{in})$, which satisfy the absorption condition (8):

$$\mathbf{r} \in \partial\Omega, \mathbf{v} \cdot \mathbf{n}_{\partial\Omega} \le 0: \quad f(\mathbf{r},\mathbf{v}) = \beta\left(v, \frac{\mathbf{v}}{v} \cdot \mathbf{n}_{\partial\Omega}\right) f(\mathbf{r}, \mathbf{v} - 2(\mathbf{v} \cdot \mathbf{n}_{\partial\Omega})\mathbf{n}_{\partial\Omega}), \quad 0 \le \beta \le 1,$$

let us name $D_0(\Xi_{in})$.

The question: is not $D_0(\Xi_{in})$ empty? - has no answer in this paper. It is tied with formulation of sufficient conditions for existence of stationary solution, discussed above. So, we guess here: it is not empty.

If $\mathbf{v} \cdot \mathbf{n}_{\partial\Omega} \le 0$, then for velocity of electron before reflection $\mathbf{v}' = \mathbf{v} - 2(\mathbf{v} \cdot \mathbf{n}_{\partial\Omega})\mathbf{n}_{\partial\Omega}$ we obtain $\mathbf{v}' \cdot \mathbf{n}_{\partial\Omega} = (\mathbf{v} - 2(\mathbf{v} \cdot \mathbf{n}_{\partial\Omega})\mathbf{n}_{\partial\Omega}) \cdot \mathbf{n}_{\partial\Omega} = -\mathbf{v} \cdot \mathbf{n}_{\partial\Omega} \ge 0$.

The proof of the theorem below is main goal of this paper:

## 3. The theorem

A solution of integral equation

$$s(\mathbf{r}) = a(\mathbf{r}) + \int_\Omega d^3 r' G_0(\mathbf{r},\mathbf{r}') s(\mathbf{r}'), \tag{9}$$

respond $s(\mathbf{r})$, where

$$G_0(\mathbf{r},\mathbf{r}') = \int d^3 v\, \omega_{ion}(v) g(\mathbf{r},\mathbf{v};\mathbf{r}',0), \tag{10}$$

$a(\mathbf{r})$ is an arbitrary function, which is integrable under the Lebesgue sense in $\Omega$, $g$ is defined with (7) and (8), at the condition of existence of generalized solution $g_1$ of auxiliary equation

$$\left(\mathbf{v} \cdot \frac{\partial}{\partial \mathbf{r}} - \frac{e}{m_e}\mathbf{E}(\mathbf{r}) \cdot \frac{\partial}{\partial \mathbf{v}} - L_{el}(\mathbf{v}) + \omega(v)\right) g_1(\mathbf{r},\mathbf{v};\mathbf{r}',\mathbf{v}') = \delta^3(\mathbf{r}-\mathbf{r}')\delta^3(\mathbf{v}-\mathbf{v}')$$

which also satisfying boundary conditions (8), exists and is unique.

## 4. The proof

### 4.1. Lemma 1
The solution of homogenious equation

$$\left(\mathbf{v} \cdot \frac{\partial}{\partial \mathbf{r}} - \frac{e}{m_e}\mathbf{E}(\mathbf{r}) \cdot \frac{\partial}{\partial \mathbf{v}} - L_{el}(\mathbf{v}) + \omega(v)\right) f(\mathbf{r},\mathbf{v}) = 0 \tag{11}$$

in domain $\Xi_{in}$ in the class of functions $D_0(\Xi_{in})$ exists and is unique: $f(\mathbf{r},\mathbf{v}) = 0$.

#### 4.1.1. The proof of lemma 1
An existence of solution is obvious.

Let be $U = \max_{\mathbf{r} \in \Omega}(-\varphi(\mathbf{r}))$, and a number $E > U$ has chosen. Let us construct the domain $\Xi_{in}^E = \Xi_{in} \cap \{\varepsilon(\mathbf{r}, \mathbf{v}) < E\}$. Multiply the equation (11) by $2f(\mathbf{r}, \mathbf{v})$ and integrate over $\Xi_{in}^E$:

$$\iint_{\Xi_{in}^E} d^3r\, d^3v\, 2f(\mathbf{r}, \mathbf{v}) \left( \mathbf{v} \cdot \frac{\partial}{\partial \mathbf{r}} - \frac{e}{m_e} \mathbf{E}(\mathbf{r}) \cdot \frac{\partial}{\partial \mathbf{v}} - L_{el}(\mathbf{v}) + \omega(v) \right) f(\mathbf{r}, \mathbf{v}) = 0.$$

Make transformations:

$$\iint_{\Xi_{in}^E} d^3r\, d^3v\, 2f(\mathbf{r}, \mathbf{v}) \left( \mathbf{v} \cdot \frac{\partial}{\partial \mathbf{r}} - \frac{e}{m_e} \mathbf{E}(\mathbf{r}) \cdot \frac{\partial}{\partial \mathbf{v}} \right) f(\mathbf{r}, \mathbf{v}) -$$

$$- \iint_{\Xi_{in}^E} d^3r\, d^3v\, 2f(\mathbf{r}, \mathbf{v}) L_{el}(\mathbf{v}) f(\mathbf{r}, \mathbf{v}) + \iint_{\Xi_{in}^E} d^3r\, d^3v\, 2f^2(\mathbf{r}, \mathbf{v}) \omega(v) = 0,$$

$$\iint_{\Xi_{in}^E} d^3r\, d^3v \left( \frac{\partial}{\partial \mathbf{r}} \cdot \mathbf{v} - \frac{e}{m_e} \frac{\partial}{\partial \mathbf{v}} \cdot \mathbf{E}(\mathbf{r}) \right) f^2(\mathbf{r}, \mathbf{v}) -$$

$$- \iint_{\Xi_{in}^E} d^3r\, d^3v \sum_{i=1}^{3} \frac{\partial}{\partial v_i} \sum_{k=1}^{3} \omega_{el}(v)(\delta_{ik} v^2 - v_i v_k) \frac{\partial}{\partial v_k} f^2(\mathbf{r}, \mathbf{v}) +$$

$$+ 2\iint_{\Xi_{in}^E} d^3r\, d^3v \sum_{i=1}^{3} \sum_{k=1}^{3} \omega_{el}(v)(\delta_{ik} v^2 - v_i v_k) \frac{\partial f}{\partial v_i}(\mathbf{r}, \mathbf{v}) \frac{\partial f}{\partial v_k}(\mathbf{r}, \mathbf{v}) +$$

$$+ 2\iint_{\Xi_{in}^E} d^3r\, d^3v\, f^2(\mathbf{r}, \mathbf{v}) \omega(v) = 0.$$

First two integrals are reduced to integration along the boundary $\partial \Xi_{in}^E$. It consists of three kinds of sections: 1) the section, which is conditioned with the boundary of spatial domain of discharge $\Gamma_{\partial\Omega} = (\partial\Omega \times R^3) \cap \overline{\Xi}_{in}^E$, 6-dimensional normal to the boundary is $N_{\partial\Omega} = (\mathbf{n}_{\partial\Omega}, \mathbf{0})$; 2) the section, which is conditioned with upper boundary of energy $\Gamma_E = \{\varepsilon(\mathbf{r}, \mathbf{v}) = E\} \cap \overline{\Xi}_{in}^E$, the normal is $N_E = \left. \left( \frac{\partial \varepsilon}{\partial \mathbf{r}}, \frac{\partial \varepsilon}{\partial \mathbf{v}} \right) \right|_{\varepsilon=E} = \left. \left( \mathbf{E}(\mathbf{r}), \frac{m_e}{e} \mathbf{v} \right) \right|_{\varepsilon=E}$; 3) the section, which is conditioned with the lower energy thresholds for inelastic processes $\Gamma_{in} = \bigcup_{p=1}^{P} \{\varepsilon(\mathbf{r}, \mathbf{v}) = \varepsilon_p\} \cap \overline{\Xi}_{in}^E$, which consists of sections of constant energy $\varepsilon(\mathbf{r}, \mathbf{v}) = \varepsilon_p$, the normal is $N_{in} = \left. \left( \mathbf{E}(\mathbf{r}), \frac{m_e}{e} \mathbf{v} \right) \right|_{\varepsilon=\varepsilon_p}$.

First integral vanishes in sections $\Gamma_E$ and $\Gamma_{in}$ on a reason of orthogonality of the phase flow to the normal of the hypersurface of constant energy. In section $\Gamma_{\partial\Omega}$ it gives the

expression

$$\iint_{\Gamma_{\partial\Omega}} d^2r\, d^3v\, \mathbf{n}_{\partial\Omega} \cdot \mathbf{v}\, f^2(\mathbf{r},\mathbf{v}) =$$

$$= \iint_{\Gamma_{\partial\Omega}\cap\{\mathbf{n}_{\partial\Omega}\cdot\mathbf{v}>0\}} d^2r\, d^3v\, \mathbf{n}_{\partial\Omega} \cdot \mathbf{v}\, f^2(\mathbf{r},\mathbf{v}) + \iint_{\Gamma_{\partial\Omega}\cap\{\mathbf{n}_{\partial\Omega}\cdot\mathbf{v}<0\}} d^2r\, d^3v\, \mathbf{n}_{\partial\Omega} \cdot \mathbf{v}\, f^2(\mathbf{r},\mathbf{v}) =$$

$$= \iint_{\Gamma_{\partial\Omega}\cap\{\mathbf{n}_{\partial\Omega}\cdot\mathbf{v}>0\}} d^2r\, d^3v\, \mathbf{n}_{\partial\Omega} \cdot \mathbf{v}\, f^2(\mathbf{r},\mathbf{v}) + \iint_{\Gamma_{\partial\Omega}\cap\{\mathbf{n}_{\partial\Omega}\cdot\mathbf{v}<0\}} d^2r\, d^3v\, \mathbf{n}_{\partial\Omega} \cdot \mathbf{v}\, \beta^2\!\left(v, \frac{\mathbf{v}}{v}\cdot\mathbf{n}_{\partial\Omega}\right) f^2(\mathbf{r},\mathbf{v} - 2(\mathbf{v}\cdot\mathbf{n}_{\partial\Omega})\mathbf{n}_{\partial\Omega}) =$$

$$= \iint_{\Gamma_{\partial\Omega}\cap\{\mathbf{n}_{\partial\Omega}\cdot\mathbf{v}>0\}} d^2r\, d^3v\, \mathbf{n}_{\partial\Omega} \cdot \mathbf{v}\left(1 - \beta^2\!\left(v, -\frac{\mathbf{v}}{v}\cdot\mathbf{n}_{\partial\Omega}\right)\right) f^2(\mathbf{r},\mathbf{v}).$$

This expression is nonnegative due to the condition of incomplete absorption (8).

Second integral vanishes in all three sections of the boundary due to orthogonality of the boundary normal to the vector of flow density (the intergrand here is a divergence of 3D-vector, which is orthogonal to $\mathbf{v}$).

So, we have

$$\iint_{\Gamma_{\partial\Omega}\cap\{\mathbf{n}_{\partial\Omega}\cdot\mathbf{v}>0\}} d^2r\, d^3v\, \mathbf{n}_{\partial\Omega} \cdot \mathbf{v}\left(1 - \beta^2\!\left(v, -\frac{\mathbf{v}}{v}\cdot\mathbf{n}_{\partial\Omega}\right)\right) f^2(\mathbf{r},\mathbf{v}) +$$

$$+ 2\iint_{\Xi_{in}^E} d^3r\, d^3v \sum_{i=1}^{3}\sum_{k=1}^{3} \omega_{el}(v)\left(\delta_{ik}v^2 - v_i v_k\right)\frac{\partial f}{\partial v_i}(\mathbf{r},\mathbf{v})\frac{\partial f}{\partial v_k}(\mathbf{r},\mathbf{v}) +$$

$$+ 2\iint_{\Xi_{in}^E} d^3r\, d^3v\, f^2(\mathbf{r},\mathbf{v})\omega(v) = 0.$$

In order that sum of three integrals with continuous nonnegative integrands be equal to zero it is necessary that every of integrands be equal to zero in all points of the domain of integration. Taking into account the condition of incomplete absorption in first integral and positiveness of average rate of elastic collisions $\omega_{el}(v)$ at nonzero velocity, we obtain

$$f^2(\mathbf{r},\mathbf{v}) = 0, \quad (\mathbf{r},\mathbf{v}) \in \Gamma_{\partial\Omega}, \quad \beta \neq 1, \tag{12}$$

$$\sum_{i=1}^{3}\sum_{k=1}^{3}\left(\delta_{ik}v^2 - v_i v_k\right)\frac{\partial f}{\partial v_i}(\mathbf{r},\mathbf{v})\frac{\partial f}{\partial v_k}(\mathbf{r},\mathbf{v}) = 0, \quad (\mathbf{r},\mathbf{v}) \in \Xi_{in}^E, \tag{13}$$

$$f^2(\mathbf{r},\mathbf{v})\omega(v) = 0, \quad (\mathbf{r},\mathbf{v}) \in \Xi_{in}^E. \tag{14}$$

The equality (12) means, that $f(\mathbf{r},\mathbf{v})$ vanishes in the section $\Gamma_{\partial\Omega}$ at any velocity (if total reflection is absent $\beta \neq 1$). The equality (13) means, that inside $\Xi_{in}^E$ the function $f(\mathbf{r},\mathbf{v})$ is not dependent of the velocity direction:

$$f(\mathbf{r},\mathbf{v}) = \tilde{f}(\mathbf{r},v^2). \tag{15}$$

The equality (14) demands $f(\mathbf{r}, \mathbf{v})$ to vanish everywhere in $\Xi_{in}^{E}$, where the average rate of inelastic processes $\omega(v) \neq 0$: $f(\mathbf{r}, \mathbf{v}) = 0$, $(\mathbf{r}, \mathbf{v}) \in \left( \Xi_{in}^{E} \setminus \{\omega(v) = 0\} \right)$. But the subset of low kinetic energies $(\mathbf{r}, \mathbf{v}) \in \left( \Xi_{in}^{E} \cap \{m_e v^2 / (2e) \leq \varepsilon_{in}\} \right)$, where $\omega(v) = 0$, is not empty and it needs additional investigation (it includes "fast" electrons having low kinetic energy, but total mechanical energy sufficient for inelastic processes). Substitution the expression (15) into equation (11) for the subset mentioned with $\omega(v) = 0$ gives

$$\left( \mathbf{v} \cdot \frac{\partial}{\partial \mathbf{r}} - \frac{e}{m_e} \mathbf{E}(\mathbf{r}) \cdot \frac{\partial}{\partial \mathbf{v}} \right) \tilde{f}(\mathbf{r}, v^2) = 0, \qquad (16)$$

or

$$\mathbf{v} \cdot \left( \frac{\partial}{\partial \mathbf{r}} \tilde{f}(\mathbf{r}, v^2) + \frac{\partial \varphi}{\partial \mathbf{r}}(\mathbf{r}) \frac{2e}{m_e} \frac{\partial}{\partial v^2} \tilde{f}(\mathbf{r}, v^2) \right) = 0.$$

1) Guess primarily, that $\mathbf{v} \neq 0$. Because the direction of the vector $\mathbf{v}$ is arbitrary, we have

$$\frac{\partial}{\partial \mathbf{r}} \tilde{f}(\mathbf{r}, v^2) + \frac{\partial \varphi}{\partial \mathbf{r}}(\mathbf{r}) \frac{2e}{m_e} \frac{\partial}{\partial v^2} \tilde{f}(\mathbf{r}, v^2) = 0. \qquad (17)$$

Let us choose some point $\mathbf{r} = \mathbf{r}_0$ in the $\Omega$ and arbitrary smooth curve in the set $\Phi(u): \mathbf{r} \in \Phi(u): \varphi(\mathbf{r}) = \varphi(\mathbf{r}_0) = -u$, which originates from this point. Because of the set $\Phi(u)$ is not necessary to be connected, not any two of points in it could be connected with curve in general. Let us name all points of the set $\Phi_1(u): \mathbf{r} \in \Phi_1(u) \subset \Phi(u)$, which one can connect by a curve with the point $\mathbf{r} = \mathbf{r}_0$, *a connected component* of the set $\Phi(u)$. At every value $u$, $0 \leq u \leq U$ we get some integer $P = P(u)$, $P = 1, 2, \ldots$ of connected components of the set $\Phi(u)$, at that $\Phi(u)$ can be presented as a sum of its non-crossing connected components:

$$\Phi(u) = \bigcup_{p=1}^{P} \Phi_p(u), \quad \Phi_p(u) \cap \Phi_q(u) = \varnothing, \ p \neq q.$$

The sets $\Phi_p(u)$, defined here, are 2D surfaces - boundaries of 3-dimensional body, which could be defined with equation $-\varphi(\mathbf{r}) \leq u$ (or equation for "additional body" $-\varphi(\mathbf{r}) > u$). If one suggests $w - \varphi(\mathbf{r}) = u$, then from inequality $w \geq 0$ it follows the equation for the body mentioned. The body is the domain of 3-dimensional space, where the electron can be arranged having total energy $u$. In presence of local minimums of potential this domain becomes many-connected, but its addendum would be unconnected set. In presence of local maximums of potential - vice versa: electron can be situated in one of "potential pits" and it cannot jump over into other "pit".

Let us fix $v^2 = \mathrm{const}$. Multiplying the equation (17) by 3D vector $\mathbf{l} = \mathbf{l}(\mathbf{r})$ tangential to the curve in $\Phi_1(-\varphi(\mathbf{r}_0))$ at given point $\mathbf{r}$, we obtain that the derivative along the curve $\mathbf{l} \cdot (\partial \tilde{f} / \partial \mathbf{r}) = 0$, therefore function $\tilde{f}(\mathbf{r}, v^2)$ has constant value in the connected component $\Phi_1(-\varphi(\mathbf{r}_0))$ of the set $\Phi(-\varphi(\mathbf{r}_0))$.

Now let us choose other curve $C : \mathbf{r} = \mathbf{c}(u)$, which issues from point $\mathbf{r} = \mathbf{r}_0$ such that electric potential $\varphi(\mathbf{r})$ be only increasing (or only decreasing) along it. The parameter of the curve can be chosen as $u = -\varphi(\mathbf{r})$. Any point $\mathbf{r}$ of this curve belongs to one of connected components of the set $\Phi(-\varphi(\mathbf{r}))$, the suffix $p$ could be regarded as fixed, for example, as equal to unit. The curve can be prolonged upward (downward) the electric potential as far as local maximum (local minimum) of the potential has reached.

Let us name a *connectedness domain of electric potential* the set

$$\Omega_c = \bigcup_{\mathbf{r} \in C} \Phi_1(-\varphi(\mathbf{r})).$$

Here the curve $C$ is meant as longest. If two of such curves connect the same pair of local maximum and local minimum of potential, they define identical set $\Omega_c$.

Indeed, suppose contrary: for such curves $C_1$ and $C_2$ exists a value $u'$: $u_{\min} < u' < u_{\max}$ of parameter, for which corresponding points of curves belong two different connected components of potential: $\mathbf{c}_1(u') \in \Phi_1(u')$, $\mathbf{c}_2(u') \in \Phi_2(u')$. If we start from point of local maximum of potential $\mathbf{r}_0 = \mathbf{c}_1(u_{\min}) = \mathbf{c}_2(u_{\min})$, the 3D spatial body $\{\mathbf{r} : u_{\min} \leq -\varphi(\mathbf{r}) \leq u'\}$ would be a connected body, having not less than two disconnected pieces of its boundary: $\{\mathbf{r} \in \Phi_1(u')\}$ and $\{\mathbf{r} \in \Phi_2(u')\}$. So, the additional set $\{\mathbf{r} : u' < -\varphi(\mathbf{r}) \leq u_{\max}\}$ is disconnected set, consisting of two insolated bodies, with boundaries $\Phi_1(u')$ and $\Phi_2(u')$. Therefore, the curve $C_1$ ends inside $\Phi_1(u')$, but the curve $C_2$ ends inside $\Phi_2(u')$, so curves could not come to a single local minimum of potential. From this contradiction it follows, that points $\mathbf{c}_1(u), \mathbf{c}_2(u)$; $u_{\min} < u < u_{\max}$ of both curves always belong the same connected component of potential.

So, a connectedness domain of electric potential is defined with pair of local minimum and local maximum of potential, and does not depend on a choice of connecting curve. It should be mentioned, that not every pair defines a connectedness domain.

Choosing arbitrary pairs, which can be connected with some curve, having monotonically changed potential along it, one can get different connectedness domains of the potential, which can cross each other, but no one can be arranged inside another entirely. The feature of a connectedness domain is to have single local minimum and single local maximum of electric potential. In the connectedness domain of potential at fixed value of the velocity the function $\tilde{f}(\mathbf{r}, v^2)$ has identical value where the potential is identical, therefore the dependence of function on coordinates takes place through the potential only:

$$\tilde{f}(\mathbf{r}, v^2) = \hat{f}\left(\varphi(\mathbf{r}), \frac{m_e v^2}{2e}\right), \quad \mathbf{r} \in \Omega_c.$$

In 6-dimensional phase space the connectedness domains of the potential originate open sets $(\Omega_c \times R^3) \cap \Xi_{in}^E$, which are connected in a sense of curves in this space (it can be proved in analogy of connectedness of $\Xi_{in}^E$ or $\Xi_{in}$). These 6-dimensional domains we name connectedness domains of the potential also. For the connectedness domain of the potential the equation (17) can be rewritten as:

$$\frac{\partial \varphi}{\partial \mathbf{r}}(\mathbf{r}) \left( \frac{\partial}{\partial \varphi} \hat{f}(\varphi, w) + \frac{\partial}{\partial w} \hat{f}(\varphi, w) \right) = 0.$$

1.1) Where electric field is not vanish, we have

$$\frac{\partial}{\partial \varphi} \hat{f}(\varphi, w) + \frac{\partial}{\partial w} \hat{f}(\varphi, w) = 0.$$

$$\hat{f}(\varphi, w) = \chi(w - \varphi),$$

here $\chi$ is any arbitrary-differentiable function. With the use of (15), in the connectedness domain of the potential at $\mathbf{v} \neq 0$ we obtain:

$$f(\mathbf{r}, \mathbf{v}) = \tilde{f}(\mathbf{r}, v^2) = \chi\left( \frac{m_e v^2}{2e} - \varphi(\mathbf{r}) \right) = \chi(\varepsilon(\mathbf{r}, \mathbf{v})). \qquad (18)$$

1.2) The set of singular points $\{\mathbf{r} : \mathbf{E}(\mathbf{r}) = 0\}$ is a closed set. Consider any its connected component. Obviously, the potential has constant value along it. The equation (17) gives $\frac{\partial}{\partial \mathbf{r}} \tilde{f}(\mathbf{r}, v^2) = 0$, which means a dependence on one scalar value $v^2$ only. It does not contradict the expression (18), therefore this expression can be extended into the set of singular points of electric potential mentioned, if such exist.

Function $f$ is a continuous function, so, if two different connectedness domains of the potential cross each other at some value of total energy $\varepsilon$, the values $\chi(\varepsilon)$ of appropriate functions are equal. The exception is a case when the set $\{\varepsilon(\mathbf{r}, \mathbf{v}) = \text{const}\}$ consists of non-connected subsets of $\Xi_{in}^E$, in this case values of $\chi(\varepsilon)$ in different subsets can differ at the same value of total energy $\varepsilon$.

Thus, all solutions of equation (11) on $\Xi_{in}^E$ at $\mathbf{v} \neq 0$ exhaust with functions

$$f(\mathbf{r},\mathbf{v}) = \chi_q(\varepsilon(\mathbf{r},\mathbf{v})), \quad \varepsilon(\mathbf{r},\mathbf{v}) = \frac{m_e \mathbf{v}^2}{2e} - \varphi(\mathbf{r}), \quad q = 1,...,Q(\varepsilon), \tag{19}$$

where $Q$ is a number of connected components of hypersurface of constant energy. This number can differ the number $P$ of connected components of the potential: $Q(u) \neq P(u), 0 \leq u \leq U$, because connected components of equal potential $-\varphi(\mathbf{r}) = u$ are divided by presence both local minimums and local maximums of the potential, but connected components of constant energy $\varepsilon(\mathbf{r},\mathbf{v}) = \text{const}$ are divided by presence of local maximums of potential only. This difference is stipulated by non-negatively defined kinetic energy. If $Q > 1$, there exists several local maximums of potential - minimums of potential energy of the electron. Obviously, the number $Q$ of connected components of hypersurface of constant energy cannot exceed the number of local maximums of potential.

For some value of total energy $\varepsilon_1$, $\varepsilon_1 > \varepsilon_{in}$ choose any of connected components of set $\{\varepsilon(\mathbf{r},\mathbf{v}) = \varepsilon_1\}$, such one, that its crossing with $\Xi_{in}^E$ is not empty (it means, that the electron, which is arranged over given "potential pit", has the kinetic energy in some place, which exceeds the threshold of inelastic processes $\varepsilon_{in}$). In the point $\mathbf{r}_0^{(q)}$ of local minimum of the electron potential energy we have $\varepsilon(\mathbf{r}_0^{(q)},\mathbf{v}) > \varepsilon_{in} - \varphi(\mathbf{r}_0^{(q)})$, or, other way to say, $m_e \mathbf{v}^2/(2e) > \varepsilon_{in}$, therefore here $\omega(v) > 0$. Because of continuity, there exists a neighborhood of the point $\mathbf{r}_0^{(q)}$ in the connected component of energy hypersurface given, in which everywhere $\omega(v) > 0$, and therefore $f(\mathbf{r},\mathbf{v}) = 0$ (in account of (14)). The union of all such neighborhoods of the point $\mathbf{r}_0^{(q)}$ bounds with the set, on which $\omega(v) = 0$, where the statement (19) is true. The continuity of $f$ at this boundary means, that $\chi_q(\varepsilon_1) = 0$. So, the statement (19) extends the lemma condition onto the set of low kinetic energies $(\mathbf{r},\mathbf{v}) \in (\Xi_{in}^E \cap \{m_e \mathbf{v}^2/(2e) \leq \varepsilon_{in}\})$, where $\omega(v) = 0$, because this set is reachable along the hypersurface of constant energy from the points of high kinetic energy. Thus, the lemma condition is true in the set $\Xi_{in}^E \setminus \{\mathbf{v} = 0\}$.

2) The set $\Xi_{in}^E \cap \{\mathbf{v} = 0\}$ is a boundary set for open set $\Xi_{in}^E \setminus \{\mathbf{v} = 0\}$. Since the solution of equation (11) is seeking on the class of continuous functions, the statement (19) can be extended on zero values of the velocity by passage to the limit, in result of which one extends the lemma condition onto whole set $\Xi_{in}^E$.

3) Since the choice of upper limit of the energy $\varepsilon(\mathbf{r}, \mathbf{v}) = \mathrm{E}$ for the set $\Xi_{in}^{\mathrm{E}}$ is not bounded from the top with any condition, for any point $(\mathbf{r}, \mathbf{v}) \in \Xi_{in}$ one can find such $\mathrm{E}$, that this point would belong to the set $\Xi_{in}^{\mathrm{E}}$ also, therefore the lemma condition is extended onto whole set $\Xi_{in}$.

The lemma is proved.

### 4.1.2 The remark

The domain of uniqueness of zero solution of homogeneous equation (11) can be extended into zone $\Xi_s$ of slow electrons along those hypersurfaces of constant energy, which cross the boundary $\partial \Xi$. It follows from the fact, that (12) at $\beta \neq 1$ one can obtain by integration over $\Xi_s$ also. Therefore the condition of absorption makes zero solution unique also in those potential pits, where absorbing walls of the device are present. So, the zone of *absence* of uniqueness remains the pits, which are insolated from the walls, on the bottom of which indefinite amount of slow electrons could present.

It is clear from physical sense, that accumulation of slow electrons in the insolated pit reduces positive spatial charge of the pit, in result of which the pit vanishes. Besides, the motion of slow electrons should satisfy another kinetic equation with non-linear integral of their mutual collisions, which leads to maxwellization of their distribution. But the consideration of back influence of the electron distribution on the electric field, also the kinetics of slow electrons, does not include into subject of this paper.

### 4.1.3 The conjugate equation

Let $f(\mathbf{r}, \mathbf{v}) \in D_0(\Xi_{in}^{\mathrm{E}})$. Multiply non-homogeneous equation

$$\left( \mathbf{v} \cdot \frac{\partial}{\partial \mathbf{r}} - \frac{e}{m_e} \mathbf{E}(\mathbf{r}) \cdot \frac{\partial}{\partial \mathbf{v}} - L_{el}(\mathbf{v}) + \omega(v) \right) f(\mathbf{r}, \mathbf{v}) = s_1(\mathbf{r}, \mathbf{v}) \qquad (20)$$

on arbitrary function $h(\mathbf{r}, \mathbf{v}) \in D(\Xi_{in}^{\mathrm{E}})$ and integrate over the domain $\Xi_{in}^{\mathrm{E}}$:

$$\iint_{\Xi_{in}^{\mathrm{E}}} d^3 r \, d^3 v \, h(\mathbf{r}, \mathbf{v}) \left( \mathbf{v} \cdot \frac{\partial}{\partial \mathbf{r}} - \frac{e}{m_e} \mathbf{E}(\mathbf{r}) \cdot \frac{\partial}{\partial \mathbf{v}} - L_{el}(\mathbf{v}) + \omega(v) \right) f(\mathbf{r}, \mathbf{v}) = \iint_{\Xi_{in}^{\mathrm{E}}} d^3 r \, d^3 v \, h(\mathbf{r}, \mathbf{v}) s_1(\mathbf{r}, \mathbf{v}).$$

Making integration by parts in the left hand side of the equation, we obtain:

$$\iint_{\Gamma_{\partial\Omega}} d^2r\, d^3v\, \mathbf{n}_{\partial\Omega} \cdot \mathbf{v}\, h(\mathbf{r},\mathbf{v}) f(\mathbf{r},\mathbf{v}) +$$

$$+ \iint_{\Xi_{in}^E} d^3r\, d^3v\, f(\mathbf{r},\mathbf{v}) \left( -\mathbf{v} \cdot \frac{\partial}{\partial \mathbf{r}} + \frac{e}{m_e} \mathbf{E}(\mathbf{r}) \cdot \frac{\partial}{\partial \mathbf{v}} - L_{el}(\mathbf{v}) + \omega(v) \right) h(\mathbf{r},\mathbf{v}) = \quad (21)$$

$$= \iint_{\Xi_{in}^E} d^3r\, d^3v\, h(\mathbf{r},\mathbf{v}) s_1(\mathbf{r},\mathbf{v}).$$

Designate:

$$D_1 = \mathbf{v} \cdot \frac{\partial}{\partial \mathbf{r}} - \frac{e}{m_e} \mathbf{E}(\mathbf{r}) \cdot \frac{\partial}{\partial \mathbf{v}} - L_{el}(\mathbf{v}) + \omega(v), \quad (22)$$

$$D_1^* = -\mathbf{v} \cdot \frac{\partial}{\partial \mathbf{r}} + \frac{e}{m_e} \mathbf{E}(\mathbf{r}) \cdot \frac{\partial}{\partial \mathbf{v}} - L_{el}(\mathbf{v}) + \omega(v). \quad (23)$$

From the obtained equality (21) it is apparently, that if function $h$ satisfies conjugate homogeneous equation

$$D_1^* h(\mathbf{r},\mathbf{v}) = 0, \quad (24)$$

also the "conjugate" condition of incomplete absorption:

$$\mathbf{r} \in \partial\Omega, \mathbf{v} \cdot \mathbf{n}_{\partial\Omega} \geq 0: \quad h(\mathbf{r},\mathbf{v}) = \beta\left(v, -\frac{\mathbf{v}}{v} \cdot \mathbf{n}_{\partial\Omega}\right) h(\mathbf{r}, \mathbf{v} - 2(\mathbf{v} \cdot \mathbf{n}_{\partial\Omega})\mathbf{n}_{\partial\Omega}), \quad 0 \leq \beta \leq 1 \quad (25)$$

(in which a sign of scalar product of the velocity on the normal is opposite to that was in (8)), then the left hand side of the equality, in account of transformations

$$\iint_{\Gamma_{\partial\Omega}} d^2r\, d^3v\, \mathbf{n}_{\partial\Omega} \cdot \mathbf{v}\, h(\mathbf{r},\mathbf{v}) f(\mathbf{r},\mathbf{v}) =$$

$$= \iint_{\Gamma_{\partial\Omega} \cap \{\mathbf{n}_{\partial\Omega} \cdot \mathbf{v} > 0\}} d^2r\, d^3v\, \mathbf{n}_{\partial\Omega} \cdot \mathbf{v}\, h(\mathbf{r},\mathbf{v}) f(\mathbf{r},\mathbf{v}) + \iint_{\Gamma_{\partial\Omega} \cap \{\mathbf{n}_{\partial\Omega} \cdot \mathbf{v} < 0\}} d^2r\, d^3v\, \mathbf{n}_{\partial\Omega} \cdot \mathbf{v}\, h(\mathbf{r},\mathbf{v}) f(\mathbf{r},\mathbf{v}) =$$

$$= \iint_{\Gamma_{\partial\Omega} \cap \{\mathbf{n}_{\partial\Omega} \cdot \mathbf{v} > 0\}} d^2r\, d^3v\, \mathbf{n}_{\partial\Omega} \cdot \mathbf{v}\, \beta\left(v, -\frac{\mathbf{v}}{v} \cdot \mathbf{n}_{\partial\Omega}\right) h(\mathbf{r}, \mathbf{v} - 2(\mathbf{v} \cdot \mathbf{n}_{\partial\Omega})\mathbf{n}_{\partial\Omega}) f(\mathbf{r},\mathbf{v}) +$$

$$+ \iint_{\Gamma_{\partial\Omega} \cap \{\mathbf{n}_{\partial\Omega} \cdot \mathbf{v} < 0\}} d^2r\, d^3v\, \mathbf{n}_{\partial\Omega} \cdot \mathbf{v}\, h(\mathbf{r},\mathbf{v}) \beta\left(v, \frac{\mathbf{v}}{v} \cdot \mathbf{n}_{\partial\Omega}\right) f(\mathbf{r}, \mathbf{v} - 2(\mathbf{v} \cdot \mathbf{n}_{\partial\Omega})\mathbf{n}_{\partial\Omega}) =$$

$$= \iint_{\Gamma_{\partial\Omega} \cap \{\mathbf{n}_{\partial\Omega} \cdot \mathbf{v} > 0\}} d^2r\, d^3v\, \mathbf{n}_{\partial\Omega} \cdot \mathbf{v}\, \beta\left(v, -\frac{\mathbf{v}}{v} \cdot \mathbf{n}_{\partial\Omega}\right) h(\mathbf{r}, \mathbf{v} - 2(\mathbf{v} \cdot \mathbf{n}_{\partial\Omega})\mathbf{n}_{\partial\Omega}) f(\mathbf{r},\mathbf{v}) -$$

$$- \iint_{\Gamma_{\partial\Omega} \cap \{\mathbf{n}_{\partial\Omega} \cdot \mathbf{v} > 0\}} d^2r\, d^3v\, \mathbf{n}_{\partial\Omega} \cdot \mathbf{v}\, h(\mathbf{r}, \mathbf{v} - 2(\mathbf{v} \cdot \mathbf{n}_{\partial\Omega})\mathbf{n}_{\partial\Omega}) \beta\left(v, -\frac{\mathbf{v}}{v} \cdot \mathbf{n}_{\partial\Omega}\right) f(\mathbf{r},\mathbf{v}) = 0,$$

vanishes, and consequently, the condition of the orthogonality of the right hand side of (20) to all solutions of the homogeneous problem (24), (25) becomes the condition of solvability of the equation (20) (the Fredholm condition). But for this problem the next lemma is true:

### 4.2 Lemma 2

The unique solution of conjugate homogeneous equation (24) at "conjugate" condition of incomplete absorption (25) on class of functions $h(\mathbf{r},\mathbf{v}) \in D(\Xi_{in}^E)$ is $h(\mathbf{r},\mathbf{v}) = 0$.

### 4.2.1 The proof

is completely analogues to the proof of the lemma 1.

### 4.3 Lemma 3

Generalized function $g_1(\mathbf{r},\mathbf{v};\mathbf{r}',\mathbf{v}')$, $(\mathbf{r},\mathbf{v};\mathbf{r}',\mathbf{v}') \in \Xi_{in} \times \Xi_{in}$, which satisfies the equations:

$$\left(\mathbf{v}\cdot\frac{\partial}{\partial \mathbf{r}} - \frac{e}{m_e}\mathbf{E}(\mathbf{r})\cdot\frac{\partial}{\partial \mathbf{v}} - L_{el}(\mathbf{v}) + \omega(v)\right)g_1(\mathbf{r},\mathbf{v};\mathbf{r}',\mathbf{v}') = \delta^3(\mathbf{r}-\mathbf{r}')\delta^3(\mathbf{v}-\mathbf{v}'),\tag{26}$$

$$\left(-\mathbf{v}'\cdot\frac{\partial}{\partial \mathbf{r}'} + \frac{e}{m_e}\mathbf{E}(\mathbf{r}')\cdot\frac{\partial}{\partial \mathbf{v}'} - L_{el}(\mathbf{v}') + \omega(v')\right)g_1(\mathbf{r},\mathbf{v};\mathbf{r}',\mathbf{v}') = \delta^3(\mathbf{r}-\mathbf{r}')\delta^3(\mathbf{v}-\mathbf{v}')\tag{27}$$

and the conditions of incomplete absorption:

$$(\mathbf{r}\in\partial\Omega, \mathbf{v}\cdot\mathbf{n}_{\partial\Omega}\leq 0):\quad g_1(\mathbf{r},\mathbf{v};\mathbf{r}',\mathbf{v}') = \beta\left(v,\frac{\mathbf{v}}{v}\cdot\mathbf{n}_{\partial\Omega}\right)g_1(\mathbf{r},\mathbf{v}-2(\mathbf{v}\cdot\mathbf{n}_{\partial\Omega})\mathbf{n}_{\partial\Omega};\mathbf{r}',\mathbf{v}'),\tag{28}$$

$$(\mathbf{r}'\in\partial\Omega, \mathbf{v}'\cdot\mathbf{n}_{\partial\Omega}\geq 0):\quad g_1(\mathbf{r},\mathbf{v};\mathbf{r}',\mathbf{v}') = \beta\left(v',-\frac{\mathbf{v}'}{v'}\cdot\mathbf{n}_{\partial\Omega}\right)g_1(\mathbf{r},\mathbf{v};\mathbf{r}',\mathbf{v}'-2(\mathbf{v}'\cdot\mathbf{n}_{\partial\Omega})\mathbf{n}_{\partial\Omega}),\tag{29}$$

if it exists, is unique. The support of the function belongs to the set

$$(\Xi_{in}\times\Xi_{in}) \cap \{\varepsilon(\mathbf{r},\mathbf{v}) = \varepsilon(\mathbf{r}',\mathbf{v}')\}.$$

Physical sense of the lemma: the electron, which appear in some point $\mathbf{r}'$ having the velocity $\mathbf{v}'$, conserves its total energy during its motion as long as it leaves the energy hypersurface in result of the act of inelastic collision or absorption with a wall.

The equation (26) and conjugate equation (27), also the conditions of incomplete absorption (28) and (29), right and "conjugate", are interrelated: one of them follows from another. Indeed, let us write down the equation (20) with arbitrary right hand side in the operator form: $D_1 f = s_1$. The equations (26) and (27) in that form take a view: $D_1 \hat{g}_1 = \hat{I}$, $D_1^* \hat{g}_1^* = \hat{I}$ (here $\hat{I}$ is identical operator, $\hat{g}_1$ and $\hat{g}_1^*$ designate linear integral operators with the kernel $g_1(\mathbf{r},\mathbf{v};\mathbf{r}',\mathbf{v}')$ and the kernel conjugated). Let us multiply (26) from the right on $s_1$, then obtain: $D_1 \hat{g}_1 s_1 = s_1$, from which it follows $f = \hat{g}_1 s_1$. Now let us multiply (20) from the left on $\hat{g}_1$ and obtain: $\hat{g}_1 D_1 f = \hat{g}_1 s_1 = f$. Because of $f$ one can consider an arbitrary function (which satisfies a condition of differentiability and the conditions of incomplete absorption in the boundary), from the expression obtained it follows $\hat{g}_1 D_1 = \hat{I}$, which is identical to (27). Consequently from (26) it follows (27). And back, from (27) it follows: $\hat{g}_1 D_1 = \hat{I}$, $\hat{g}_1 D_1 f = f$, and then with account of (20): $\hat{g}_1 s_1 = f$, $D_1 \hat{g}_1 s_1 = D_1 f = s_1$, from which, because of arbitrariness of $s_1$, we obtain $D_1 \hat{g}_1 = \hat{I}$.

### 4.3.1 The proof

*Existence and uniqueness.* The equation (26) is equivalent to the operator equation $D_1 \hat{g}_1 = \hat{I}$, and the equation (27) - the operator equation $D_1^* \hat{g}_1^* = \hat{I}$, or, that is the same, to the equation $\hat{g}_1 D_1 = \hat{I}$; in the aggregate they are equivalent to existence of the reciprocal operator $\hat{g}_1 = D_1^{-1}$, or to existence and uniqueness classical (not generalized) solution of the equation (20) at an arbitrary continuous right hand side. For existence of a solution of non-homogeneous equation (20) at the boundary conditions of incomplete absorption it is necessary the orthogonality of the right hand side of the equation to all non-zero solutions of conjugate homogeneous equation (24) at "conjugate" condition of incomplete absorption (25). By lemma 2 it was proved, that such solutions do not exist, thus right hand side of (20) has no additional restrictions. As mentioned above, the formulation of *sufficient* conditions of existence of solution of the equation (20) is not an easy problem, so we do not consider it here.

For uniqueness of solution of non-homogeneous equation (20) it is necessary and sufficient, that the homogeneous equation (11) would have unique solution (equal to zero). The last was proved by lemma 1.

*The support of $g_1$.* The proof of the lemma 1 (also the lemma 2) remains true in that case, if from the set $\Xi_{in}$ one excludes any layer $\{\varepsilon_1 \le \varepsilon(\mathbf{r}, \mathbf{v}) \le \varepsilon_2\}$. At that, it is not necessary to put any boundary condition in the boundary of the layer, because the vector field of phase flow is tangential to the boundary in any point, and the flux through the boundary from appropriate integral terms vanishes. In result, the equalities (13) and (14) become true again, and the proof of lemma 1 and lemma 2 for the set $\Xi_{in} \setminus \{\varepsilon_1 \le \varepsilon(\mathbf{r}, \mathbf{v}) \le \varepsilon_2\}$ is carried out in analogy to made above.

Let

$$\eta(\mathbf{r}) = \begin{cases} C_\eta \exp\left(\dfrac{1}{r^2 - 1}\right), & |\mathbf{r}| < 1; \\ 0, & |\mathbf{r}| \ge 1. \end{cases} \quad C_\eta = \left(\int_0^1 4\pi r^2 \exp\left(\dfrac{1}{r^2 - 1}\right) dr\right)^{-1}$$

V.V.Gorin. *Uniqueness theorem for the non-local ionization source in glow discharge and hollow cathode*be the normalized "cap" [15]. Build a sequence of "caps", which are convergent to the delta-function of right hand side of (26) or (27):

$$\eta_n(\mathbf{r},\mathbf{v};\mathbf{r}',\mathbf{v}') = n^6 \eta(n(\mathbf{r}-\mathbf{r}'))\eta(n(\mathbf{v}-\mathbf{v}')).$$

(Here the convergence treats in a sense of weak convergence of linear functionals, see [16].) The support of any "cap" is concentrated in the right product of balls

$$\operatorname{supp}\eta_n = \{|\mathbf{r}-\mathbf{r}'|\leq 1/n\}\times\{|\mathbf{v}-\mathbf{v}'|\leq 1/n\}.$$

At fixed $\mathbf{r}', \mathbf{v}'$ consider the equation $D_1 f_n = \eta_n$ respond the unknown $f_n$. Because of continuity of the function $\varepsilon(\mathbf{r},\mathbf{v})$ on its variables, for any $\delta\varepsilon > 0$ one can find such number $n_0$, to begin from which it is true the estimation for the energy:

$$|\varepsilon(\mathbf{r},\mathbf{v})-\varepsilon(\mathbf{r}',\mathbf{v}')| < \delta\varepsilon, \quad (\mathbf{r},\mathbf{v})\in\operatorname{supp}\eta_n, \quad n\geq n_0.$$

If now one chooses $\varepsilon_1 = \varepsilon(\mathbf{r}',\mathbf{v}')-\delta\varepsilon$, $\varepsilon_2 = \varepsilon(\mathbf{r}',\mathbf{v}')+\delta\varepsilon$, than we obtain that the equation $D_1 f_n = \eta_n$ on the set $\Xi_{in}\setminus\{\varepsilon_1 \leq \varepsilon(\mathbf{r},\mathbf{v})\leq \varepsilon_2\}$ is homogeneous: $D_1 f_n = 0$. Because of lemma 1 it has here single solution - zero. Hence, the support of the solution of the equation $D_1 f_n = \eta_n$ belongs to the set $\Xi_{in}\cap\{\varepsilon_1 \leq \varepsilon(\mathbf{r},\mathbf{v})\leq \varepsilon_2\}$. Because $\delta\varepsilon$ could be chosen arbitrarily small, the support of (generalized) function $f = \lim_{n\to\infty} f_n$ (weak limit) belongs to the crossing of all such sets, namely, to the set $\Xi_{in}\cap\{\varepsilon(\mathbf{r},\mathbf{v})=\varepsilon(\mathbf{r}',\mathbf{v}')\}$ (at fixed $\mathbf{r}',\mathbf{v}'$). Now if one fixes $\mathbf{r},\mathbf{v}$ and consider in analogy the conjugate equation $D_1^* h_n = \eta_n$, one, using lemma 2, comes to analogues result respond variables $\mathbf{r}',\mathbf{v}'$. Therefore the support of generalized function $g_1$ belongs to the set $(\Xi_{in}\times\Xi_{in})\cap\{\varepsilon(\mathbf{r},\mathbf{v})=\varepsilon(\mathbf{r}',\mathbf{v}')\}$.

The lemma is proved.

**4.4 Lemma 4**

The solution $g(\mathbf{r},\mathbf{v};\mathbf{r}',\mathbf{v}')$ of the equation (7)

$$\left(\mathbf{v}\cdot\frac{\partial}{\partial\mathbf{r}} - \frac{e}{m_e}\mathbf{E}(\mathbf{r})\cdot\frac{\partial}{\partial\mathbf{v}} - L(\mathbf{v})\right)g = \delta^3(\mathbf{r}-\mathbf{r}')\delta^3(\mathbf{v}-\mathbf{v}')$$

in the class of generalized functions on the set $(\mathbf{r},\mathbf{v};\mathbf{r}',\mathbf{v}')\in\Xi_{in}\times\Xi_{in}$, at the boundary conditions of incomplete absorption (8), in a condition of existence of the function $g_1(\mathbf{r},\mathbf{v};\mathbf{r}',\mathbf{v}')$ (see lemma 3), exists, is unique and has a property:

$$g(\mathbf{r},\mathbf{v};\mathbf{r}',\mathbf{v}') = 0 \text{ at } \varepsilon(\mathbf{r},\mathbf{v}) > \varepsilon(\mathbf{r}',\mathbf{v}'). \tag{30}$$

Physical sense of inequality (30): the electron, which has arisen in glow discharge in the phase point $\mathbf{r}', \mathbf{v}'$ at the energy $\varepsilon(\mathbf{r}', \mathbf{v}')$, can not gain more energy in the process of motion under elastic and inelastic scattering on "motionless" atoms of gas, because inelastic processes make its mechanics of motion dissipative.

### 4.4.1 The proof

Rewrite the equation (7) in the form

$$\left( \mathbf{v} \cdot \frac{\partial}{\partial \mathbf{r}} - \frac{e}{m_e} \mathbf{E}(\mathbf{r}) \cdot \frac{\partial}{\partial \mathbf{v}} - L_{el}(\mathbf{v}) + \omega(\mathbf{v}) \right) g(\mathbf{r}, \mathbf{v}; \mathbf{r}', \mathbf{v}') =$$
$$= \int d^3 v'' \mu(\mathbf{v}, \mathbf{v}'') g(\mathbf{r}, \mathbf{v}''; \mathbf{r}', \mathbf{v}') + \delta^3(\mathbf{r} - \mathbf{r}') \delta^3(\mathbf{v} - \mathbf{v}').$$

Because of property (5), if a point $(\mathbf{r}, \mathbf{v}) \in \Xi_{in}$, then point $(\mathbf{r}, \mathbf{v}'') \in \Xi_{in}$ (otherwise the factor $\mu$ vanishes). Using the superposition principle [15] p. 195, let us represent the solution of equation (7) as a sum of solutions with elementary sources from (26). It enables to represent the equation (7) in integral form

$$g(\mathbf{r}, \mathbf{v}; \mathbf{r}', \mathbf{v}') = \iint_{\Xi_{in}} d^3 r''' d^3 v''' g_1(\mathbf{r}, \mathbf{v}; \mathbf{r}''', \mathbf{v}''') \left( \int d^3 v'' \mu(\mathbf{v}''', \mathbf{v}'') g(\mathbf{r}''', \mathbf{v}''; \mathbf{r}', \mathbf{v}') + \delta^3(\mathbf{r}''' - \mathbf{r}') \delta^3(\mathbf{v}''' - \mathbf{v}') \right),$$

or

$$g(\mathbf{r}, \mathbf{v}; \mathbf{r}', \mathbf{v}') = \iint_{\Xi_{in}} d^3 r''' d^3 v''' g_1(\mathbf{r}, \mathbf{v}; \mathbf{r}''', \mathbf{v}''') \int d^3 v'' \mu(\mathbf{v}''', \mathbf{v}'') g(\mathbf{r}''', \mathbf{v}''; \mathbf{r}', \mathbf{v}') + g_1(\mathbf{r}, \mathbf{v}; \mathbf{r}', \mathbf{v}').$$

Because of property (5) one can change the order of integration by the velocities $\mathbf{v}''', \mathbf{v}''$:

$$g(\mathbf{r}, \mathbf{v}; \mathbf{r}', \mathbf{v}') = \iint_{\Xi_{in}} d^3 r''' d^3 v'' \int d^3 v''' g_1(\mathbf{r}, \mathbf{v}; \mathbf{r}''', \mathbf{v}''') \mu(\mathbf{v}''', \mathbf{v}'') g(\mathbf{r}''', \mathbf{v}''; \mathbf{r}', \mathbf{v}') + g_1(\mathbf{r}, \mathbf{v}; \mathbf{r}', \mathbf{v}').$$

Due to (5) the integration here is made over set

$$(\mathbf{r}'', \mathbf{v}'''; \mathbf{v}'') \in \Xi_{in} \times \left( R^3 \cap \left\{ \frac{m_e v'''^2}{2e} < \frac{m_e v''^2}{2e} - \varepsilon_{in} \right\} \right).$$

From structure of this set we have inequality: $v''^2 > v'''^2 + \frac{2e}{m_e} \varepsilon_{in} > v'''^2$, so, if $(\mathbf{r}'', \mathbf{v}''') \in \Xi_{in}$ than $(\mathbf{r}'', \mathbf{v}'') \in \Xi_{in}$ also. This make possible to change the order of integration mentioned.

We obtained the integral equation

$$g(\mathbf{r}, \mathbf{v}; \mathbf{r}', \mathbf{v}') = \iint_{\Xi_{in}} d^3 r'' d^3 v'' K(\mathbf{r}, \mathbf{v}; \mathbf{r}'', \mathbf{v}'') g(\mathbf{r}'', \mathbf{v}''; \mathbf{r}', \mathbf{v}') + g_1(\mathbf{r}, \mathbf{v}; \mathbf{r}', \mathbf{v}'), \tag{31}$$

where

$$K(\mathbf{r}, \mathbf{v}; \mathbf{r}'', \mathbf{v}'') = \int d^3 v''' g_1(\mathbf{r}, \mathbf{v}; \mathbf{r}'', \mathbf{v}''') \mu(\mathbf{v}''', \mathbf{v}''). \tag{32}$$

Using (5), we obtain:

$$K(\mathbf{r},\mathbf{v};\mathbf{r}'',\mathbf{v}'') = \int_{\frac{m_e \mathbf{v}'''^2}{2e} < \frac{m_e \mathbf{v}''^2}{2e} - \varepsilon_{in}} d^3v''' g_1(\mathbf{r},\mathbf{v};\mathbf{r}'',\mathbf{v}''') \mu(\mathbf{v}''',\mathbf{v}'')$$

From the lemma 3 for the function $g_1$ and from the property of energy dissipation (5) it follows that the support of function $K$ (by variables $\mathbf{r},\mathbf{v}$) can be arranged only at

$$\varepsilon(\mathbf{r},\mathbf{v}) = \varepsilon(\mathbf{r}'',\mathbf{v}''') < \varepsilon(\mathbf{r}'',\mathbf{v}'') - \varepsilon_{in},$$

from which we have:

$$K(\mathbf{r},\mathbf{v};\mathbf{r}',\mathbf{v}') = 0 \text{ при } \varepsilon(\mathbf{r},\mathbf{v}) \geq \varepsilon(\mathbf{r}',\mathbf{v}') - \varepsilon_{in}. \tag{33}$$

Physical sense: the operator

$$\hat{K} f(\mathbf{r},\mathbf{v}) = \iint_{\Xi_{in}} d^3r' d^3v' K(\mathbf{r},\mathbf{v};\mathbf{r}',\mathbf{v}') f(\mathbf{r}',\mathbf{v}') \tag{34}$$

reduces the energy of electrons not less then the value $\varepsilon_{in}$ of minimal energy of all inelastic processes: if function $f(\mathbf{r},\mathbf{v})$ has a support restricted by energy, which belongs to the set $\Xi_{in} \cap \{\varepsilon(\mathbf{r},\mathbf{v}) < \varepsilon_1\}$, then the support of the function $\hat{K} f(\mathbf{r},\mathbf{v})$ must belong to the set $\Xi_{in} \cap \{\varepsilon(\mathbf{r},\mathbf{v}) < \varepsilon_1 - \varepsilon_{in}\}$.

Let $I_{in}(\mathbf{r},\mathbf{v}) = 1, (\mathbf{r},\mathbf{v}) \in \Xi_{in}$, $I_{in}(\mathbf{r},\mathbf{v}) = 0, (\mathbf{r},\mathbf{v}) \notin \Xi_{in}$ be an indicator of the set $\Xi_{in}$. The operator $\hat{K} : D_0(\Xi_{in}) \Rightarrow D_0(\Xi_{in})$ one can present as a product of two operators: $\hat{K} = \hat{g}_1 \hat{\mu}, \quad \hat{\mu} : D_0(\Xi_{in}) \Rightarrow C(\Xi_{in}), \quad \hat{g}_1 : C(\Xi_{in}) \Rightarrow D_0(\Xi_{in})$, where

$$\hat{\mu} f(\mathbf{r},\mathbf{v}) = I_{in}(\mathbf{r},\mathbf{v}) \int_\Omega d^3r' \delta^3(\mathbf{r}-\mathbf{r}') \int d^3v' \mu(\mathbf{v},\mathbf{v}') f(\mathbf{r}',\mathbf{v}').$$

The indicator of set $\Xi_{in}$ is included to avoid consideration of zone $\Xi_s$ of slow electrons (the process of transformation "fast" electrons into slow ones is not interesting for the problem stated). Because of the function $f \in D_0(\Xi_{in})$ is finite, its support is restricted by energy with the value $\varepsilon(\mathbf{r},\mathbf{v}) < \varepsilon_1$, therefore the function $s_2(\mathbf{r},\mathbf{v}) = \hat{\mu} f(\mathbf{r},\mathbf{v})$ is finite also, for its support the inequality $\varepsilon(\mathbf{r},\mathbf{v}) < \varepsilon_1 - \varepsilon_{in}$ is true.

The operator $\hat{K}$ acts into the same manifold of functions, on which it is defined, so there exist powers of the operator $\hat{K}^2 = \hat{K}\hat{K}, \quad \hat{K}^3 = \hat{K}\hat{K}\hat{K},...$ and so on.

In the operator form (where the names of integral operators correspond to its kernels) the equation (31) takes form: $\hat{g} = \hat{K}\hat{g} + \hat{g}_1$, or $\left(\hat{I} - \hat{K}\right)\hat{g} = \hat{g}_1$. Its formal solution is the Neumann series:

$$\hat{g} = \left(\hat{I} - \hat{K}\right)^{-1}\hat{g}_1 = \left(\sum_{n=0}^{\infty}\hat{K}^n\right)\hat{g}_1.$$

However, because the energy is restricted from the bottom, but every action of the operator $\hat{K}$ reduces the energy of electron not less then $\varepsilon_{in}$, so an operator $\hat{K}^n\hat{g}_1$ vanishes on functions of the class $C(\Xi_{in}^E)$ at $n > E/\varepsilon_{in}$, other to say, the operator $\hat{K}$ is nilpotent on functions, which support is restricted from energy top with energy $E$, the index of nilpotency is equal to $n = [E/\varepsilon_{in}] + 1$ (here the brackets designate integral part of number: [3.14] = 3). Thus the series has finite number of summands:

$$\hat{g} = \left(\sum_{n=0}^{N}\hat{K}^n\right)\hat{g}_1, \quad N = [\varepsilon(\mathbf{r}', \mathbf{v}')/\varepsilon_{in}]. \tag{35}$$

Existence and uniqueness of $g(\mathbf{r}, \mathbf{v}; \mathbf{r}', \mathbf{v}')$ are consequences of those for kernels $g_1$ and $K$, which are included into formula for the operator, and unambiguity of finite number of operations in it. Since there is no summands, which increase initial energy of electron $\varepsilon(\mathbf{r}', \mathbf{v}')$ (which electron had before first act of inelastic scattering), then $g(\mathbf{r}, \mathbf{v}; \mathbf{r}', \mathbf{v}') = 0$ at $\varepsilon(\mathbf{r}, \mathbf{v}) > \varepsilon(\mathbf{r}', \mathbf{v}')$.

Because of arbitrary choice of upper boundary of energy in limits of the domain of "fast" electrons the statement of the lemma can be extended into whole set $(\mathbf{r}, \mathbf{v}; \mathbf{r}', \mathbf{v}') \in \Xi_{in} \times \Xi_{in}$.

The lemma is proved.

**4.4.2 Normalized spaces**

In manifolds $D_0(\Xi_{in})$ and $C(\Xi_{in})$ one can introduce the norm of space $L_1$ in such way:

$$s \in C(\Xi_{in}): \quad \|s\| = \int_{\Xi_{in}} d^3r\, d^3v\, |s(\mathbf{r}, \mathbf{v})|,$$

$$f \in D_0(\Xi_{in}): \quad \|f\| = \|D_1 f\|, \quad D_1 f \in C(\Xi_{in}).$$

Such definition of the norm is convenient in the sense, that the norms of the operator $D_1$ and its reverse operator $\hat{g}_1$ are equal to unit, and they turn out to be restricted and continuous automatically. These norms have physical sense: they are equal to the total flow of "fast" electrons from non-negative distributed source in all volume of V.V.Gorin. *Uniqueness theorem for the non-local ionization source in glow discharge and hollow cathode*discharge. In this way, the norm of the operator $\hat{\mu}$ is equal to the norm of the operator $\hat{K} = \hat{g}_1 \hat{\mu}$ and it does not exceed unit, because of conservation of flow of electrons, which undergo inelastic scattering. In sequence, the operators $\hat{\mu}$ and $\hat{K}$ are restricted and continuous also.

However, the other norm $\|f\| = \int_{\Xi_{in}} d^3r d^3v |f(\mathbf{r},\mathbf{v})|$ could have physical sense for the space $D_0(\Xi_{in})$ also, at non-negative $f$ it coincides with (average) total number of "fast" electrons in the discharge volume. But if such norm to use, the operator $\hat{g}_1$ can be unrestricted on $C(\Xi_{in})$. Indeed, let an electron potential energy "pit" exists in the discharge volume, which has a bottom $\varepsilon = \varepsilon_{\min}$, a depth $\delta\varphi_p > \varepsilon_{in}$ and does not border with walls of discharge volume. Then for equation (20) there is a solution

$$f(\mathbf{r},\mathbf{v}) = \chi_p(\varepsilon(\mathbf{r},\mathbf{v})), \quad \operatorname{supp}\chi_p \in \{\varepsilon_{in} \leq (\varepsilon - \varepsilon_{\min}) \leq \delta\varphi_p\}, \quad D_1 f = \omega(v)\chi_p(\varepsilon(\mathbf{r},\mathbf{v})) = s_1(\mathbf{r},\mathbf{v}).$$

Obviously, one can build a sequence of functions

$$\chi_p^{(n)}(\varepsilon) = \eta_1\left(\frac{(n+1)(\varepsilon - \varepsilon_{\min}) - n\varepsilon_{in} - \delta\varphi_p}{\delta\varphi_p - \varepsilon_{in}}\right),$$

$$\eta_1(x) = C_\eta \exp\left(\frac{1}{x^2-1}\right), -1 < x < 1; \quad \eta_1(x) = 0, x \geq 1; \quad C_\eta = \left(\int_{-1}^{1} dx \exp\left(\frac{1}{x^2-1}\right)\right)^{-1}.$$

which weakly converges to $\delta(\varepsilon - \varepsilon_{\min} - \varepsilon_{in})$; at that way the norm "number of particles" would converge to finite positive limit, but in the same time the flow $\|s_1^{(n)}\|$ of the electrons, which undergo inelastic scattering, would tend to zero because of factor $\omega(v)$. If now one builds new sequence from previous one, in which every term of initial sequence is divided on the number $\|s_1^{(n)}\|$, then now for every term of new sequence the flow is equal to unit, and in the same time "the number of particles" would increase unrestrictedly.

Nevertheless, in the norm "number of particles" for functions from $D_0$ the operator $\hat{g}_1$ would be restricted on manifolds $C(\Xi_{in+\delta\varepsilon})$, $\delta\varepsilon > 0$, and, in particular, on $C(\Xi_{ion})$. Unrestrictedness of $\hat{g}_1$ on the whole $C(\Xi_{in})$ is tied with increase of concentration of "fast" electrons in bringing near to zone of slow electrons, because of necessity of the flow conservation in a stationary problem, though the rate $\omega(v)$ of inelastic processes tends to zero in reaching the boundary of division of "fast" and slow electrons. As mentioned above, the gaining of electrons in the pit must lead to "deformation" of the pit itself and to formation the configuration of electric field of discharge, which does not enable too high concentrations of electrons.

**4.5 Lemma 5**

The operator $\hat{G}_0$: $s_1(\mathbf{r}) = \int_\Omega d^3r' G_0(\mathbf{r},\mathbf{r}') s(\mathbf{r}')$ is defined identically and has properties:

a) *Shrinking of a support*: if a support of non-negative function $s(\mathbf{r})$ belongs to the set $\overline{\Omega}_u = \overline{\Omega} \cap \{-\varphi(\mathbf{r}) \leq u\}$, then the support of the non-negative function $s_1(\mathbf{r}) = \hat{G}_0 s(\mathbf{r})$ belongs to the set

$$\overline{\Omega}_{u-\varepsilon_{ion}} = \overline{\Omega} \cap \{-\varphi(\mathbf{r}) \leq u - \varepsilon_{ion}\},$$

and the enclosure $\Omega_{u-\varepsilon_{ion}} \subset \Omega_u$ is true.

b) *Nilpotency*: since the potential energy is restricted from the top and bottom, there exists positive integer power $M$ of the operator $\hat{G}_0$, which turns any function $s(\mathbf{r})$, $\mathbf{r} \in \Omega$, integrable on $\Omega$, identically into zero: $\left(\hat{G}_0\right)^M s(\mathbf{r}) = 0$.

**4.5.1 The proof**

*Unambiguity*. On the definition

$$G_0(\mathbf{r},\mathbf{r}') = \int d^3v\, \omega_{ion}(v) g(\mathbf{r},\mathbf{v};\mathbf{r}',0).$$

Though unambiguity of definition of the function $g(\mathbf{r},\mathbf{v};\mathbf{r}',0)$ at fixed $\mathbf{r}'$ is guaranteed by lemma 4 only in the domain $(\mathbf{r},\mathbf{v}) \in \Xi_{in}$ for "fast" electrons, the integration by velocities in zone $\Xi_s$ of slow electrons does not contribute, because first factor in the integral, the rate of ionization $\omega_{ion}(v)$, is equal to zero in zone $\Xi_s$ (because of (1)). Therefore function $G_0(\mathbf{r},\mathbf{r}')$ is defined identically.

a) From lemma 4 it follows, that $g(\mathbf{r},\mathbf{v};\mathbf{r}',0) = 0$ at $\frac{m_e v^2}{2e} - \varphi(\mathbf{r}) > -\varphi(\mathbf{r}')$, other to say, non-zero values of $g$ are possible only at $\frac{m_e v^2}{2e} \leq \varphi(\mathbf{r}) - \varphi(\mathbf{r}')$. But, due to property (1), to provide the factor $\omega_{ion}(v)$ not to vanish it is necessary, that the condition $\varepsilon_{ion} < \frac{m_e v^2}{2e}$ be true, from which follows $\varepsilon_{ion} < \frac{m_e v^2}{2e} \leq \varphi(\mathbf{r}) - \varphi(\mathbf{r}')$, or $-\varphi(\mathbf{r}) < -\varphi(\mathbf{r}') - \varepsilon_{ion}$. It means that

$$G_0(\mathbf{r},\mathbf{r}') = 0 \text{ at } -\varphi(\mathbf{r}) \geq -\varphi(\mathbf{r}') - \varepsilon_{ion}. \tag{36}$$

Since the support of the function $s(\mathbf{r}')$ belongs to the set $\overline{\Omega}_u = \overline{\Omega} \cap \{-\varphi(\mathbf{r}') \leq u\}$, than from (36) it follows, that non-zero values of $s_1(\mathbf{r})$ are possible only at $-\varphi(\mathbf{r}) < -\varphi(\mathbf{r}') - \varepsilon_{ion} < u - \varepsilon_{ion}$, other to say, the support of the function $s_1(\mathbf{r})$ belongs to $\overline{\Omega}_{u-\varepsilon_{ion}}$. The enclosure $\Omega_{u-\varepsilon_{ion}} \subset \Omega_u$ is obvious because of positive value of ionization threshold $\varepsilon_{ion}$.

The statement a) is proved.

b) Since the electric potential in a gas discharge device is restricted from top and bottom, statement b) is true at $M = [U/\varepsilon_{ion}] + 1$, where $U = \max_{\mathbf{r} \in \Omega}\{-\varphi(\mathbf{r})\}$.

The lemma is proved.

### 4.6 The proof of the theorem

The equation (9) in operator form takes a view: $s(\mathbf{r}) = a(\mathbf{r}) + \hat{G}_0\, s(\mathbf{r})$, or $\left(\hat{I} - \hat{G}_0\right) s(\mathbf{r}) = a(\mathbf{r})$.

Formal solution is given by the Neumann series:

$$s(\mathbf{r}) = \left(\hat{I} - \hat{G}_0\right)^{-1} a(\mathbf{r}) = \left(\sum_{m=0}^{\infty} \hat{G}_0^{\,m}\right) a(\mathbf{r}).$$

However, because of nilpotency of the operator $\hat{G}_0$, which was proved in the lemma 5, this series has only finite number of summands:

$$s(\mathbf{r}) = \left(\sum_{m=0}^{M} \hat{G}_0^{\,m}\right) a(\mathbf{r}), \quad M = [U/\varepsilon_{ion}]. \tag{37}$$

The formula (37) obtained defines identically the solution of the equation (9), and thus it proves an existence of the solution (at the condition of existence of solution the equation (26)) and its uniqueness.

The theorem is proved.

### 5. Discussion of results and conclusions

In the proof of the uniqueness theorem for solution of equation (9) there are built many of useful constructions and formulae. All of them have clear physical sense and enable to understand a structure of non-local electron avalanche, which forms configuration of non-local source of ionization in glow discharge and hollow cathode.

First of all, one manages to understand, that an elementary source of "fast" electrons, which can be regarded as localized in some point of 6D phase space, supplies electrons into hypersurface of constant energy, which the source point belongs to. The elastic collisions with atoms do not change significantly the electron energy because of slow velocity of atoms in comparison with typical velocities of "fast" electrons and small mass of electrons, so elastic scattering can be regarded as conserving the electron energy. Thus, the auxiliary differential equation (26) with appropriate boundary conditions defines the distribution $g_1$ of electrons from a point source, which did not suffer inelastic scattering and wall absorption as yet. This stationary solution might exist, if a point source of electrons is balanced with distributed electron losses in inelastic scattering and wall absorption. The distribution $g_1$ has monochromatic energy spectrum (everywhere here we mean total mechanical energy $\varepsilon(\mathbf{r},\mathbf{v}) = \frac{m_e v^2}{2e} - \varphi(\mathbf{r})$).

The inelastic losses of energy generate a source of electrons having lower energy than primary ones had, this process is described with operation $s_1 = \hat{\mu} g_1$. The distribution $g_2$ of lower energy electrons gives operation $g_2 = \hat{K} g_1$. If distribution $g_2$ describes "fast" electrons, they are able to loose its energy again: $g_3 = \hat{K} g_2$, and so on.

If one guesses, that only some finite number of energy transformations are significant among all inelastic processes (for example, excitation of atom from ground state to first exited level, and its ionization from ground state to the bottom of electron continuous energy spectrum), the operator $\hat{\mu}$ would consist of finite number of monochromatic summands, having as factors the delta-functions of appropriate energy changes. Then primary point source generates discrete spectrum of energies for inelastically scattered electrons. The generalized function $g$ represents an electron distribution, which is generated by the point source ($g$ does not include the distribution of ionization product - secondary electrons, born in ionization). The formula (35) illustrates the structure of $g$.

Other hand, among inelastic processes there are present ionization processes, which generate secondary electrons - product of ionization. Primary energy of secondary electron (in assumption of its slow starting velocity, which we neglect) depends not on energy level of primary electron, but on value of electric potential in the spatial point of their birth. The potential changes continuously along hypersurface of constant energy for primary electron, so secondary electrons, starting with zero velocity, would have continuous spectrum of energy. If secondary electron is "fast", it generates its own discrete spectrum, but taken they all together, their spectrum would be continuous. The equation in the introduction of the paper (obtained in [12] by V.V. Gorin) represents the source of ionization $s(\mathbf{r})$ as a sum of contribution of electrons from the cathode, and contribution of secondary electrons from discharge volume. First are primarily monochromatic, after inelastic action they are distributed on some discrete spectrum, so the first summand has discrete energy spectrum. In second summand we see the integration inside discharge volume over $\mathbf{r}'$. As potential energy of the electron changes continuously in this integration, and secondary electrons supposed to be born with negligible kinetic energy, electron energy in a source $s(\mathbf{r}')$ changes continuously, and, though the kernel $G_0(\mathbf{r},\mathbf{r}')$ at fixed $\mathbf{r}'$ would give discrete spectrum (see definition (10)), the averaging on $\mathbf{r}'$ gives continuous energy spectrum of ionization with secondary electrons. So, the non-local ionization source has both discrete and continuous energy spectrum, that was found earlier by other authors [18]. Formula for distribution function, obtained in [12]:

$$f_e(\mathbf{r},\mathbf{v}) = \int_{\partial\Omega} d^2 r' g(\mathbf{r},\mathbf{v};\mathbf{r}',0) j_n(\mathbf{r}') + \int_{\Omega} d^3 r' g(\mathbf{r},\mathbf{v};\mathbf{r}',0) s(\mathbf{r}')$$

has analogues structure: first summand, from cathode electrons, has discrete energy spectrum due to constant value of potential along the cathode and discrete spectrum of the point source; second summand has continuous spectrum due to averaging in the integral over all discharge volume.

Final formula (37) for a source of ionization demonstrates a structure of non-local avalanche: operators $\hat{G}_0^m$, $m = 1,...,M$. increase primary ionization source. Other words, one primary electron ionizes in average some number of atoms, some of secondary electrons are "fast", and they, in their turn, ionize some number of atoms in addition, and so on, - as long as no "fast" electrons appear (at $m = M$). Mathematically it means: the norm of operators $\left\|\hat{G}_0^m\right\| > 0$, $m = 1,...,M$, so, the norm of operator

$$\left\|\sum_{m=0}^{M} \hat{G}_0^m\right\| = \sum_{m=0}^{M} \left\|\hat{G}_0^m\right\| = 1 + \sum_{m=1}^{M} \left\|\hat{G}_0^m\right\| > 1, \quad M = [U/\varepsilon_{ion}].$$

Here general inequality for norms is substituted with equality, because all operator summands are non-negative and act on non-negative source functions. When "more than unit" (> 1) can be substituted with "much more than unit" (>> 1), the primary ionization source (first summand of series in (37) at $m = 0$) could be very small in comparison with all sum, this is named "avalanche". As distinct from the Townsend local avalanche, where increase of number of electrons occur along one of spatial coordinates in the direction opposite to electric field, the increase in the non-local avalanche is not spatial, but rather energetic one. It is a reason why the light of hollow cathode looks like it is very uniform.

## 6. Acknowledgments

I thank Prof. A.P. Yurachkovsky for creative discussions of the paper, in result of which I managed to increase an austerity of statements and quality of results obtained. Thanks also merit the efforts of my friends and colleagues Dr. O.S. Boordo and Prof. Yu.V. Yakovenko, which were aimed to make paper more clear.

The work is made under support of the grant number 5499 of The Ministry of Science of Russian Federation.

## THE LIST OF REFERENCES

1. *Paschen F.* // Ann der Phys. – 1916. - V. 50, P. 901.
2. *Engel A., Shteenbeck M.* Physics and technology of electric discharge in gases 2. - translation from German to Russian under N.A. Kaptsov edition / Moscow: ONTI, 1936.
3. *Yu.P. Rayzer,* Physics of gas discharge / Moscow: Nauka, 1987.
4. *B.I. Moskalev,* Discharge with hollow cathode / Moscow: Energia, 1969.
5. *Kutasi K., Donko Z.* Hybrid model of a plane-parallel hollow-cathode discharge // J. Phys. D: Appl. Phys. – 2000. - V. 33, P. 1081 – 1089.
6. *Sigeneger F., Winkler R.* Study of the electron kinetics in cylindrical hollow cathodes by a multi-term approach // Eur. Phys. J. AP. – 2002. - V. 19, P. 211 – 223.
7. *Baguer N., Bogaerts A., Gijbels R.* Hollow cathode glow discharge in He: Monte Carlo-Fluid model combined with a transport model for metastable atoms // J. Applied Physics. – 2003. - V. 93, N. 1, P. 47 – 55.
8. *Sigeneger F., Donko Z., Loffhagen D.* Boltzmann equation and particle-fluid hybrid modelling of a hollow cathode discharge // Eur. Phys. J. Appl. Phys. – 2007. - V. 38, P. 161-167.
9. *Derzsi A., Hartmann P., Korolov I., Karacsony J., Bano G. Donko Z.* On the accuracy

and limitations of fluid models of the cathode region of dc glow discharges // J. Phys. D: Appl. Phys. – 2009. - V. 42, 225204.

10. *Hilbert D.* Grundzüge einer allgemeinen Theorie der linearen Integralgleichungen. - B. G. Teubner, Leipzig, 1912.

11. *Gorin V.V.* Non-local source equation for linear stationary kinetic equation // International conference "Differential equations and topology". - 2008. Moscow, Russia, P. 43 - 44.

12. *Gorin V.V.* Non-local model of hollow cathode and glow discharge – theory calculations and experiment comparison // Eur. Phys. J. D. – 2010. – V. 59, P. 241–247, DOI: 10.1140/epjd/e2010-00165-9.

13. *Gorin V.V.* A Mathematical Model of Plane Glow Discharge and Hollow Cathode Effect // Ukr. J. Phys. – 2008. - V. 53, N. 4, P. 366-372.

14. *Gorin V.V.* Integral equation for source of ionization in hollow cathode // Preprint of Cornell Univ. Lib.: http://arxiv.org  0902.2655.

15. *V.S. Vladimirov,* The equations of mathematical physics / Moscow: Nauka, 1971.

16. *A.N. Kolmogorov, S.V. Fomin,* Elements of function theory and functional analysis / Moscow: Nauka, 1976.

17. *L.V. Kantorovich, G.P. Akilov*, Functional analysis / Moscow: Nauka, 1984.

18. *J.P. Boeuf and E. Marode* // J. Phys. D: Appl. Phys. - 1982, V. 15, P. 2169 doi:10.1088/0022-3727/15/11/012